# On Demographic Transformation: Why We Need to Think Beyond Silos

Nicholle Mae Amor Tan Maravilla[1] and Myles Joshua Toledo Tan[1,2,3,4,5,6,7,*]

[1] Yo-Vivo Corporation, Bacolod City, Negros Occidental, 6100, Philippines
[2] Department of Electrical and Computer Engineering, Herbert Wertheim College of Engineering, University of Florida, Gainesville, Florida, 32611, United States of America
[3] Department of Epidemiology, College of Public Health & Health Professions and College of Medicine, University of Florida, Gainesville, Florida, 32610, United States of America
[4] Biology Program, College of Arts and Sciences, University of St. La Salle, Bacolod City, Negros Occidental, 6100, Philippines
[5] Department of Natural Sciences, College of Arts and Sciences, University of St. La Salle, Bacolod City, Negros Occidental, 6100, Philippines
[6] Department of Chemical Engineering, College of Engineering and Technology, University of St. La Salle, Bacolod City, Negros Occidental, 6100, Philippines
[7] Department of Electronics Engineering, College of Engineering and Technology, University of St. La Salle, Bacolod City, Negros Occidental, 6100, Philippines

* Correspondence:
Myles Joshua Toledo Tan, mylesjoshua.tan@medicine.ufl.edu

## I. Introduction

Many developed countries are experiencing a remarkable demographic shift, characterized by an aging population and rapidly declining birth rates. According to the World Health Organization (WHO), as of 2020, the global population aged 60 and above outnumbered children under the age of 5. It has also been predicted that in the year 2050, the fraction of the world's population over 60 years of age will increase from 12 to 22 percent, almost double (WHO, 2024a). This drastic change in demographics of the human population creates biological, economic, and social challenges, particularly in the areas of geriatric care, healthcare systems, and workforce sustainability (Sander et al, 2014). For instance, countries like Japan, Korea, Germany, and Italy are heavily affected by the decline of working-age populations resulting in underfunded social systems, tight labor markets, and overstretched medical and care sectors (World Economic Forum, 2021).

Alarmed by declining population growth rates, many governments in developed countries are implementing strategic measures to mitigate the resulting socioeconomic impacts. For instance, nations such as Japan, China, and South Korea have launched matchmaking initiatives aimed at encouraging marriage and childbearing (Chai, 2021; Watts, 2002; Lee, 2008). In parallel, rising numbers of older adults have prompted the enactment of policies that promote healthy, active aging and expand access to cost-effective, long-term healthcare interventions (Cylus et al., 2019). However, as aging populations continue to increase and working-age populations decline, healthcare systems face mounting challenges. These include a growing shortage of healthcare providers and the increasing complexity of geriatric care (Jones & Dolsten, 2024).

To address these challenges, many technology companies and researchers are actively developing artificial intelligence (AI) and robotics to enhance healthcare support for older adults. AI technologies such as machine learning (ML), natural language processing (NLP), and predictive analytics offer transformative tools to support clinical decision-making, personalize treatment plans, enable remote monitoring, and deliver a range of essential care services for aging populations (Shiwani et al., 2023).

Furthermore, the integration of AI into healthcare systems can help mitigate persistent and emerging issues, including shortages of healthcare professionals and the increasing complexity and volume of medical data (Hazarika, 2020). To ensure effective implementation, policymakers should work closely with healthcare providers, AI developers, and older adults to co-design innovative, inclusive, and responsive solutions.

However, recent policy efforts have disproportionately focused on addressing declining birth rates, often overlooking the equally urgent challenge of a rapidly aging population. Striking a balance between the needs of younger and older generations is essential for building a sustainable and inclusive society. As we confront the twin demographic pressures of falling fertility and increasing life expectancy, it is imperative to examine their root causes, associated risks, and long-term societal implications. To effectively respond to these complex and interconnected issues, we propose a transdisciplinary approach, one that integrates insights from diverse fields to co-create holistic, innovative, and enduring solutions.

## II. The Decline in Birth Rates: What is Driving the Change?

*Socio-economic Factors*

Developed countries tend to have lower birth rates than middle- and low-income countries, largely because of higher levels of human capital. Human capital refers to individuals' education, skills, knowledge, and overall well-being, all of which significantly contribute to economic productivity. In the context of this paper, human capital plays a central role in shaping how people perceive and make decisions about fertility, which in turn influences broader demographic trends (Wang & Sun, 2016). According to Roser (2014), women in today's generation are more highly educated and play a significant role in the working-age population. This means that well-educated and professional women often face higher opportunity costs, which makes them less likely to have more children. In addition, urbanization contributes to declining birth rates because of the rising costs of raising children. The financial pressures of urban living, such as housing, healthcare, and education, make it more difficult for families to support larger households. As people pursue career development and enjoy greater personal freedom in urbanized environments, they often view smaller families as more financially manageable. (White et al, 2008).

*Socio-cultural Factors*

Traditionally, religion was a significant factor influencing household childbearing decisions in developed countries. It played a vital role in shaping family values and social expectations. Families more actively engaged in religious practices often expressed a desire for larger families, leading to higher birth rates. However, as societies become more modernized, many individuals move away from traditional religious beliefs and practices, and tend to prefer having fewer children. (Adsera, 2006). In China, where filial piety remains a core family value, many young, marriageable individuals in urban areas feel overwhelmed or discouraged by the idea of marriage, largely due to career pressures and growing influences from Western culture (Chai, 2021).

Over time, the idealization of family has shifted. Today, many societies view smaller, more manageable family sizes as practical. In addition, working adults are increasingly focused on career advancement and personal development rather than starting families, often delaying marriage and parenthood as a result (Furstenberg, 2019). People living in urban areas also have greater access to modern contraceptives, which provides them with more control over their reproductive choices (White et al., 2008).

Abortion is legal in most developed countries and is recognized as a fundamental human right by millions of people around the world. However, abortion laws vary widely. Twenty-four countries ban it entirely, while most industrialized nations permit it. About 100 countries impose restrictions, often for health-related or fetal concerns. In some cases, vague legal definitions of fetal impairment create uncertainty for medical professionals (Center for Reproductive Rights, 2024). In countries where abortion is fully legal, whether restricted or not, this may contribute to lower birth rates. In contrast, in countries heavily influenced by religious beliefs, such as the Philippines where abortion is illegal, the ban has contributed to higher birth rates (Perez et al., 2022).

*Health and Lifestyle Factors*

As a result of unhealthy lifestyles, medication use, exposure to chemicals or radiation, age, and other hereditary or health-related conditions, individuals may develop infertility. According to the World Health Organization (2024), around 17.5 percent, or about one in every six people of reproductive age, experience infertility worldwide. In men, infertility is commonly caused by dysfunctional semen ejection due to obstruction of the reproductive tract, varicoceles, or abnormal sperm function or quality. In women, it is usually caused by tubal and uterine disorders, polycystic ovarian syndrome (PCOS), and hormonal imbalances.

One of the leading causes of infertility among women today is PCOS, a chronic and incurable condition that results in hormonal imbalances, irregular menstrual cycles, elevated androgen levels, diabetes, and cystic ovaries. Approximately 8 to 13 percent of reproductive-aged women worldwide have PCOS, and 70 percent of affected women remain undiagnosed (WHO, 2023). Furthermore, 56.6 percent of women with PCOS experience mental health challenges. Among them, 40 percent have depression and 16.6 percent have mood disorders (The Lancet Regional Health – Europe, 2022). Depression and other mood disorders, whether related to PCOS or not, often reduce the desire to have children (Golovina et al., 2023). Promoting mental health and overall well-being is therefore essential for improving fertility rates.

Despite these challenges, many fertility care services are available today, including in vitro fertilization (IVF) and other types of medically assisted reproduction (WHO, 2024b). Governments in developed countries such as Japan and China have begun allocating funds to support fertility treatment in efforts to increase birth rates (Watts, 2002; Huang et al., 2024). These initiatives reflect growing recognition of the need to address population decline by helping families overcome fertility-related difficulties.

### III. Matchmaking and Reproduction Policies

In countries experiencing significant declines in birth rates, governments are making urgent decisions to protect their economies. For example, Japan's population is projected to decline by nearly half by the year 2050. This poses a serious threat to its economy, as the labor force is expected to shrink by 4.5 million by 2025. At the same time, increasing allocations will be required for pensions and healthcare services for older adults. These demographic shifts contribute to economic imbalances by reducing productivity, straining social welfare systems, and increasing the financial burden of long-term care. In response, Japan has invested billions of yen in fertility treatments and matchmaking events to encourage childbirth and marriage (Watts, 2002).

China is facing a similar challenge. Its population decreased by two million in 2023, and this trend is expected to continue due to persistently low fertility rates (Council on Foreign Relations, 2024). To counteract this, the Chinese government has allocated funds to support reproductive services and matchmaking programs (Chai, 2021). Other countries such as the Republic of Korea, Singapore, Poland, Hungary, Belarus, Turkey, and Russia have also adopted pronatalist policies aimed at increasing birth rates. These include financial incentives to encourage marriage and larger families. For instance, Singapore offers "baby bonuses," Hungary provides interest-free loans to couples with three children, and Russia grants "maternity capital" to mothers with a second or third child (Gietel-Basten et al., 2022).

However, while these policies aim to address underpopulation, they may also pose risks by restricting women's reproductive freedom. Reproduction is a fundamental human right that supports personal autonomy, freedom, and equality. The ability to choose when and how many children to have is essential to individual liberty, especially for women, and is protected under international human rights frameworks (Center for Reproductive Rights, 2003). Additionally, pronatalist and matchmaking policies could lead to unintended consequences for future generations. Policymakers must carefully consider the long-term impacts of such strategies. The historical example of China's One Child Policy, which reduced the population by an estimated 250 million people, shows how a population control measure, while addressing one problem, can contribute to future demographic challenges (Kane and Choi, 1999). If modern pronatalist efforts are implemented without adequate planning, they may result in social imbalances, increased pressure on women, or population growth that outpaces infrastructure development.

## IV. Navigating the Impact of Longer Life Expectancy on Society and Healthcare

While matchmaking and pronatalist policies remain a priority for many governments, it is equally important to address the broader challenges that accompany declining birth rates, particularly the growing need to care for an expanding aging population.

Advancements in medical care and technology, public health interventions, improved living conditions, and other socio-economic factors have contributed to a significant increase in life expectancy (Olshansky et al., 2024). Globally, life expectancy rose from 34 years in 1913 to over 70 years by 2022 (World Economic Forum, 2023). This trend has continued in 2024 and is projected to reach 85 years by the year 2100 (Statista, 2024). However, these gains in longevity have created a demographic shift in which the aging population is increasing more rapidly than the younger population, largely due to declining mortality and fertility rates, especially in high-income countries (Harper, 2019). As older adults begin to outnumber younger age groups, the pressure on social and healthcare systems will intensify, requiring substantial adjustments to meet their growing needs.

Biologically, aging leads to molecular and cellular damage that gradually reduces physical and cognitive function (WHO, 2024a). This reality demands increased healthcare attention for older adults, many of whom remain in the workforce instead of entering retirement (Scott, 2021). Despite this, healthcare systems in developed countries are not fully equipped to meet the complex needs of aging populations. Challenges include shortages in the healthcare workforce and insufficient responses to key social determinants of health such as income, housing, transportation, education, and social support (Jones & Dolsten, 2024). To address these gaps, many developed nations are recruiting healthcare professionals from middle- and low-income countries, offering better salaries and employment opportunities. This trend can benefit individual workers who seek improved quality of life, but it also creates challenges for the source countries, which experience shortages in their own healthcare systems (Aluttis et al., 2014).

For example, many Filipino healthcare providers are pursuing opportunities in countries such as the United States, Canada, Australia, and the United Kingdom. While higher wages are a strong motivator, their decisions are also influenced by a desire for better living conditions and access to quality healthcare and education (Maravilla & Tan, 2021). However, this migration trend creates a delicate balance between global labor mobility and local healthcare needs. As a result, many middle- and low-income countries are simultaneously losing skilled professionals while attracting retirees from high-income countries who seek affordable healthcare and lower living costs (Holecki et al., 2020). If this trend continues, it supports projections that by 2050, 80 percent of the world's aging population could reside in middle- and low-income countries (World Economic Forum, 2023).

The migration of retirees from high-income to lower-income countries could strain social systems and local economies, highlighting the urgent need to address underlying causes and mitigate potential consequences (Holecki et al., 2020). Policymakers must also confront challenges in delivering quality healthcare to older adults. Compared to other age groups, elderly patients often require more resources and support, face higher rates of chronic illness and disability, and encounter significant ethical dilemmas in care. There is also a recognized lack of scientific evidence to guide effective preventive and clinical interventions tailored to the aging population (National Academies Press, 2001). As both aging and migration trends evolve, a comprehensive and adaptive approach is essential to sustain the health of older populations while maintaining economic stability.

## V. The Adoption of AI and Robotics in Transforming Healthcare for Aging Populations

*Opportunities*

AI technologies offer a wide range of applications in healthcare for aging populations. First, AI tools can support clinical decision-making by improving diagnostic accuracy through clinical decision support systems (CDSS). Second, natural language processing (NLP) and speech analysis can identify trends in spoken or written language, which is particularly valuable for analyzing large volumes of data in electronic health records. Third, AI can be used to analyze medical imaging data such as X-rays, CT scans, and other diagnostic tests like ECG and EEG, aiding clinicians in detecting abnormalities and improving interpretation accuracy. Fourth, AI-driven robotic systems can aid in physical rehabilitation and physiotherapy. Some are also designed to provide therapeutic engagement and support emotional well-being. Fifth, AI-powered wearable devices can support emergency response and health monitoring by detecting movement changes or abnormal physiological signals, such as during a fall or sudden health event (Shiwani et al., 2023; Padhan et al., 2023; Alowais et al., 2023).

Beyond these general applications, the integration of sophisticated foundation models like GPT and BERT can analyze vast amounts of unstructured data from electronic health records, while techniques like multi-omics integration and radiomics can create a comprehensive molecular and imaging profile of a patient. This level of analysis requires a new type of expert within the healthcare team: the data scientist. Acting as a core member of a transdisciplinary team, the data scientist can synthesize these complex datasets to provide actionable, evidence-based insights, helping to personalize care plans for older adults with complex conditions like cancer (Tan et al., 2025b).

Despite these promising applications, continuous innovation and thoughtful implementation are necessary to ensure that AI technologies are accessible, effective, and capable of meeting the growing healthcare needs of aging populations (Alowais et al., 2023). As a result, interest in AI for healthcare has continued to increase in recent years.

According to the Organization for Economic Cooperation and Development (OECD) Artificial Intelligence Policy Observatory, more than 800 initiatives, strategies, and plans related to AI have already been published by industries and non-governmental organizations. Between 2016 and 2019 alone, tens of thousands of scientific papers and patents on AI in healthcare were produced (Castonguay et al., 2024). Additionally, a study by Tran et al. (2019) found that global collaboration in AI research for health and medicine has been extensive. Major contributions have come from countries such as the United States, Canada, several European nations, Australia, and Asian countries including China, Japan, Taiwan, and South Korea. These developments indicate a growing global commitment to designing AI-powered healthcare technologies that meet the needs of older individuals.

In the United States, a global leader in artificial intelligence, numerous AI technologies have been developed to support the needs of the aging population. One example is CarePredict[1], a wearable device worn on the dominant hand that uses AI to detect specific activities and gestures for remote monitoring. Communities that have implemented this technology report a 40 percent reduction in hospitalizations and a 60 percent reduction in falls annually. Another example is Sensi.AI[2], an AI-powered platform used in both hospitals and home care settings. This tool supports health monitoring, early detection of medical issues, caregiver assistance, and promotes greater independence among older adults.

Japan, known for its expertise in industrial robotics and humanoid robot research, has developed several robotic solutions aimed at addressing workforce shortages in elder care[3]. Examples include Robear[4], a lifting robot that assists with patient transfers; Paro[5], a therapeutic robot designed to resemble a baby seal and to provide therapeutic interaction resembling animal-assisted therapy; and Pepper[6], a robot used to facilitate social interaction and recreational exercises. These robots have attracted international attention, especially in countries that also face shortages in healthcare personnel and long-term care resources.

In Europe, the Socrates Project[7] received over €3 million in funding from the European Union to develop a robotic system that enhances emotional support and social interaction for older adults (European Commission, 2018). Meanwhile, China has also made significant advancements, becoming one of the first countries to integrate humanoid robots[8], artificial intelligence, and brain-computer interfaces (BCI) into geriatric care. These technologies are intended to improve quality of life and meet the complex needs of an aging population.

---

[1] https://www.carepredict.com/, accessed on July 3rd, 2025
[2] https://www.sensi.ai/product/, accessed on July 3rd, 2025
[3] https://www.technologyreview.com/2023/01/09/1065135/japan-automating-eldercare-robots/, accessed on July 3rd, 2025
[4] https://www.riken.jp/en/news_pubs/research_news/pr/2015/20150223_2/, accessed on July 3rd, 2025
[5] http://www.parorobots.com/, accessed on July 3rd, 2025
[6] https://www.softbank.jp/en/robot/, accessed on July 3rd, 2025
[7] http://www.socrates-project.eu/, accessed on July 3rd, 2025
[8] http://www.china.org.cn/china/2025-01/08/content_117651421.htm, accessed on July 3rd, 2025

However, to ensure the full effectiveness of these innovations, developers and policymakers must carefully address the technical, social, and ethical challenges associated with integrating AI and robotics into elder care. These include concerns related to privacy, safety, emotional well-being, human oversight, and equitable access.

*Challenges and Ethical Considerations*

Despite significant efforts to develop AI technologies for elder care, several concerns continue to surround their integration and adoption. One major issue is the ethical implications of using AI, particularly with respect to protecting patient confidentiality and data privacy. Gaining the trust of patients and healthcare providers depends on how securely and transparently these technologies manage sensitive information (Badawy & Shaban, 2025). For example, Spangler et al. (2022) found that older adults expressed privacy concerns when using voice assistant systems like Amazon Alexa and Google Assistant, especially due to the perceived risk of identity and financial theft. Although legal frameworks exist to protect data, they are ineffective if not properly enforced. The UK National Health Service, for instance, violated data protection laws by sharing patient data with Google DeepMind, raising serious ethical and legal concerns (Shiwani et al., 2023). Such lapses may result in legal consequences, identity theft, data breaches, and a loss of public trust. Organizations could also face reputational damage and regulatory sanctions. Therefore, the responsible use of AI in healthcare must be guided by ethical principles, including deontological ethics, which evaluates actions based on adherence to moral duties (Kant, 1785), and consequentialism, which judges actions by their outcomes (Mill, 1861). Adopting these ethical frameworks can help promote a more accountable and trustworthy implementation of AI technologies (Tan & Maravilla, 2024).

Another essential concern is the need for compassion in the delivery of care to older adults, especially those experiencing suffering. In the development of AI technologies, empathy and compassion are often overlooked due to economic pressures, the demands of an aging population, and cost-saving policies. This tension between compassion and efficiency has resulted in negative consequences, including moral distress and burnout among healthcare professionals who struggle to balance patient-centered care with productivity goals (Kerasidou, 2020). While AI can assist in enhancing clinical outcomes, the human touch remains essential. Physicians offer comfort, interpret subtle behavioral cues, and build therapeutic relationships in ways machines cannot (Cordero, 2024). Since AI lacks emotional intelligence and interpersonal sensitivity, it is vital to incorporate compassionate care principles into its design. Morrow et al. (2023) proposed the human-AI system of intelligent caring, which includes six elements: (1) awareness of suffering, (2) understanding of suffering, (3) connection with the person suffering, (4) judgment about how to respond, (5) intention to alleviate suffering, and (6) reflection on the impact of the response. Applying this framework could help ensure that AI tools support rather than replace compassionate care in geriatric settings.

Ensuring accuracy in health data analysis and treatment personalization is also critical for building trust in AI systems. Inaccurate predictions in diagnostics or treatment planning can lead to misdiagnoses or harmful dosing errors. For example, a pilot study by Bernstein et al. (2023) demonstrated that while AI can support radiologists in identifying disease, it is still prone to false positives and false negatives that may negatively affect patient outcomes. Reliable and continuous data collection can improve the performance of AI models. If older adults use wearable devices and advanced biosensors for real-time monitoring, they may benefit from earlier detection of health issues, better management of chronic conditions, and more timely interventions (Tan et al., 2025a). However, the successful implementation of such tools depends on user acceptance. Many older individuals are reluctant to adopt new digital health technologies, often due to mistrust or unfamiliarity (Ezeudoka & Fan, 2024). To address this, AI systems should be rigorously validated, regularly updated, and continuously monitored by medical professionals to ensure reliability and safety.

Finally, the cost and maintenance of AI technologies pose additional challenges. Implementing these systems often requires substantial investment in infrastructure, software, and personnel training. Ongoing expenses related to maintenance, upgrades, and regulatory compliance can further limit access, particularly in resource-constrained settings.

Beyond the cost of the technology itself, successful implementation is hindered by practical barriers such as the lack of interoperability with legacy Electronic Health Record (EHR) systems and the need for significant upfront investment in computational infrastructure and specialized IT staff (Hinostroza Fuentes et al., 2025). Furthermore, a significant but often overlooked barrier is the professional and epistemic gap between data scientists and clinicians, which can lead to communication challenges and misaligned priorities if not actively managed within a transdisciplinary framework (Tan et al., 2025b).

Although countries like Japan have developed AI-powered robotic systems to support elder care, these technologies have not eliminated labor costs. In fact, they may require additional personnel to operate, maintain, and supervise the robots (Wright, 2023). Therefore, it is important to develop affordable AI solutions that do not compromise care quality. Both public and private sectors should invest in accessible technologies while ensuring that healthcare systems retain a strong focus on human dignity, empathy, and patient-centered care.

## VI.     Transdisciplinary Collaboration

Transdisciplinary collaboration emphasizes the co-creation of knowledge through the active involvement of not only researchers from different disciplines, but also stakeholders, practitioners, and community members. This approach fosters innovation and more effectively addresses complex, real-world challenges by bridging the gaps between theory and practice, science and society, and academia and industry (Jacobi et al., 2022). In contrast, multidisciplinary and interdisciplinary collaborations, while valuable, generally operate within more bounded disciplinary frameworks. Multidisciplinary collaboration brings together experts from various fields who work in parallel, each contributing their domain-specific knowledge without integrating approaches (Lyu, Huang, & Liu, 2024). Interdisciplinary collaboration, on the other hand, seeks to integrate methods and theoretical frameworks from different disciplines to develop a more holistic understanding of a problem (Specht & Crowston, 2022).

Professionals across diverse sectors often engage in multidisciplinary and interdisciplinary collaborations to address demographic shifts. Engineers, software developers, and AI researchers, for example, have developed technologies such as telemedicine platforms, e-health applications, wearable biosensors, and robotic caregiving tools to meet the healthcare needs of older adults (Davenport & Kalakota, 2019). Healthcare professionals from different fields, including physicians, nurses, caregivers, physical and occupational therapists, pharmacists, and psychologists, collaborate to improve the delivery of geriatric care (Keijsers et al., 2016). Similarly, addressing infertility often involves coordinated care from professionals in reproductive medicine, psychology, and social work (Peterson et al., 2012). On the population level, sociologists, economists, demographers, and policymakers work together to formulate strategies for sustainable demographic development (Kugler & Rhamey, 2023).

However, while multidisciplinary and interdisciplinary approaches yield valuable contributions, they may fall short in addressing the deeply interconnected and systemic nature of demographic challenges. These complex issues often require the integrated thinking, shared understanding, and co-production of knowledge that transdisciplinary collaboration uniquely offers. For example, Galvin et al. (2014) examined collaborative care models for Alzheimer's disease and dementia, which emphasize team-based, patient-centered strategies involving both professionals and family caregivers. These models rely on leadership support, evidence-based practices, appropriate technology, and active engagement with patients and their families, which are key transdisciplinary elements that foster trust, communication, and improved outcomes. Such approaches have been associated with enhanced quality of care, better adherence to clinical guidelines, reduced behavioral symptoms, and improved patient well-being.

The momentum behind transdisciplinary collaboration is growing in a range of domains, including education, sustainability, urban planning, and environmental science. McClure et al. (2024) identified key success factors in transdisciplinary partnerships between researchers and non-academic stakeholders in African cities to address climate risks: fostering mutual understanding, creating inclusive spaces for collaboration, and valuing local motivations. Informal and formal engagement allowed researchers to reflect on their assumptions and refine their work in light of community priorities. In these collaborations, local actors emerged as "champions" who bridged academic and non-academic worlds.

Likewise, agroecology has evolved into a transdisciplinary approach to food system sustainability. Initially focused at the farm level, it now encompasses ecological, social, technological, economic, and political dimensions across the entire food value chain, from production to consumption. It aims to optimize interactions among system components while promoting environmental stewardship, social equity, and informed food choices (Garcia-Oliveira et al., 2022). These examples demonstrate that transdisciplinary collaboration can, and should, be extended to address demographic transformations by bringing together academic experts and non-academic actors such as policymakers, industry leaders, local communities, and individuals directly affected by demographic shifts.

As AI technologies continue to expand across healthcare, fostering transdisciplinary collaboration becomes even more essential, particularly for addressing the social determinants of health. While many AI development efforts take a multidisciplinary approach, a transdisciplinary strategy that includes diverse expertise, community perspectives, and lived experiences can lead to more ethical, equitable, and context-sensitive AI tools (Tan & Benos, 2025). This is especially important in areas like elder care and infertility, where the insights of those directly impacted are vital for producing relevant and responsible solutions (Rigolot, 2020).

Nonetheless, significant barriers to transdisciplinary collaboration remain. Lawless et al. (2025) observed that while many teams aspire to work transdisciplinarily, limited understanding of what it entails, along with weak support for inclusive participation, hampers its adoption. Sargent et al. (2022) outlined key institutional and interpersonal obstacles to transdisciplinary work: organizational politics, disconnects between academia and local communities, misaligned goals, poor knowledge-sharing practices, unresolved conflicts, and academic reward systems that prioritize individual achievement over team-based contributions. These structural challenges must be addressed to unlock the full potential of transdisciplinary research and implementation.

## VII. Addressing Global Demographic Shifts: A Transdisciplinary Framework for Sustainable Solutions

Many developed countries are undergoing a profound demographic transformation, characterized by a rapidly aging population and persistently low birth rates. This dual trend creates significant biological, economic, and social pressures, straining geriatric care, healthcare systems, and workforce sustainability. Countries such as Japan, South Korea, Germany, and Italy are already experiencing serious effects from shrinking working-age populations, including underfunded social systems, limited labor supply, and overwhelmed healthcare and caregiving sectors. While governments have begun implementing various strategies to mitigate these consequences, a more comprehensive and integrated approach is urgently needed. This paper proposes a transdisciplinary framework as a viable pathway for co-creating holistic and sustainable solutions to these complex demographic challenges.

*Proposed Framework for Transdisciplinary Collaboration*

The intricate and interconnected nature of contemporary demographic shifts calls for an approach that goes beyond traditional disciplinary boundaries. Declining birth rates, for instance, are not simply economic or social issues; they are shaped by a complex web of health-related factors, lifestyle choices, and cultural values. Likewise, caring for an aging population is not limited to the domain of medicine but extends into biological, economic, social, and technological considerations. While multidisciplinary and interdisciplinary collaborations offer valuable perspectives, they often fall short of the deep integration and shared understanding needed for sustainable, long-term solutions. Transdisciplinary collaboration, by actively involving non-academic stakeholders, including those directly affected by these challenges, is uniquely positioned to develop solutions that are contextually relevant, equitable, and genuinely transformative.

As this paper has outlined, the complexity of demographic issues, such as the interplay of socioeconomic, cultural, and health factors influencing birth rates, or the layered pressures on healthcare systems due to population aging, requires more than the parallel contributions of separate disciplines. Solutions must arise from genuine co-creation and integration of knowledge. The multi-causal nature of low fertility rates, driven by factors like human capital, urbanization costs, religious trends, career priorities, infertility, mental health, and abortion laws, highlights the limitations of fragmented responses from economists, sociologists, or health experts working in isolation. A singular focus, such as promoting marriage, will likely fall short if not accompanied by efforts to reduce housing costs or address career-related barriers. Therefore, the complexity of the problem demands a solution of equal complexity. Transdisciplinarity, with its emphasis on knowledge co-creation among academics, practitioners, and communities, offers a powerful framework for developing holistic responses that move beyond the simple aggregation of disciplinary insights toward a true synthesis.

*Key Principles and Components of the Framework*

A robust transdisciplinary framework for addressing demographic shifts must rest on several core principles and components:

1. **Shared Problem Definition and Goal Alignment:** The framework must begin with all participants, academic and non-academic alike, collaboratively defining the problem based on their varied perspectives. This collective understanding is essential to avoid ***team goal misalignment*** within the team. For example, addressing declining birth rates would require input from demographers, economists, sociologists, public health experts, and, most importantly, individuals and families directly affected by fertility decisions. Together, they can identify root causes and align on desired outcomes.

2. **Inclusive Stakeholder Engagement:** The framework requires the meaningful involvement of a wide range of stakeholders, including healthcare providers, social workers, policymakers, AI developers, and community members such as older adults and those facing infertility. Their lived experiences and insights are crucial for ensuring that research and solutions are practical, relevant, and responsive to real-world needs.

3. **Iterative Co-creation and Knowledge Integration:** Rather than relying on isolated contributions from each field, the framework promotes an ongoing, interactive process of dialogue and knowledge integration. This includes both structured and informal exchanges. For example, in the development of AI for elder care, continuous feedback from clinicians and older adults is necessary to help developers create tools that are accessible, responsive, and capable of meeting evolving healthcare needs. This process also plays a key role in improving accuracy in health data detection and ensuring that treatment plans are truly personalized.

4. **Ethical Integration and Trust-Building:** Ethical concerns must be addressed from the beginning of any initiative. This includes patient privacy, compassionate care, and the responsible use of personal data. Ethicists, legal professionals, and patient advocates should work closely with technologists to ensure that systems comply with ethical frameworks such as Deontological Ethics (which evaluates actions based on rules) and Consequentialism (which assesses outcomes). Building public trust, especially among older adults who may be wary of new technologies, requires transparency, inclusivity, and demonstrable benefits.

5. **Capacity Building and Mutual Learning:** The framework should foster a reciprocal learning environment where all participants gain a deeper understanding of other disciplines' perspectives and methodologies. This helps bridge *university-community disconnects* and improves *knowledge sharing*. Non-academic participants, with their practical experience and community connections, can emerge as "vital 'champions' who bridged divides and fostered teamwork," which facilitates broader adoption and impact.

6. **Flexible Governance and Adaptive Management:** Recognizing that *organizational politics* and *conflict resolution issues* can hinder collaboration, the framework calls for clear communication channels, established conflict resolution mechanisms, and strong leadership support. Policies should be regularly reviewed and adjusted to explicitly favor and reward *team achievements* rather than only individual *achievements*, thereby encouraging genuine collaboration.

*Addressing Barriers within the Framework*

The proposed transdisciplinary framework actively seeks to mitigate the commonly identified barriers to effective collaboration:

1. **Limited Insight into Team Dynamics:** To address the lack of understanding regarding how team members view, support, and engage in collaboration, the framework recommends regular reflection sessions and embedded qualitative research within teams. This continuous feedback loop enables adaptive adjustments to team structure and processes, ultimately fostering a more cohesive and effective collaborative environment.

2. **Organizational Politics & University-Community Disconnects:** To overcome internal institutional politics and the disconnect between academic institutions and community needs, the framework calls for the establishment of formal memoranda of understanding (MOUs) or partnership agreements. These agreements outline roles, responsibilities, and shared objectives, ensuring institutional accountability. Additionally, creating neutral collaborative spaces, whether physical or virtual and equally accessible to all participants, encourages interaction outside traditional organizational hierarchies.

3. **Team Goal Misalignment:** To address differing objectives among team members, the framework prioritizes participatory methods such as shared visioning workshops at the project's outset. These sessions allow all stakeholders to co-develop and commit to overarching goals and targeted outcomes. The agreed-upon goals are then periodically reviewed and reaffirmed throughout the project lifecycle.
4. **Poor Knowledge Sharing:** To improve the flow of information, the framework recommends the use of shared digital platforms for document storage, communication, and data sharing. Scheduled inter-group meetings, cross-disciplinary training sessions, and the pursuit of joint publications promote knowledge transfer and integration. Communication practices are intentionally designed to ensure complex concepts are translated into accessible language for all team members.
5. **Conflict Resolution Issues:** To proactively address disagreements within the team, the framework includes the establishment of clear protocols for conflict resolution. A neutral facilitator or mediator may be designated to guide the process. Emphasis is placed on open dialogue, empathetic listening, and a shared commitment to collaborative rather than adversarial outcomes.
6. **Policies Favoring Individual Achievements:** To shift incentives toward collaboration, the framework advocates for policy reforms at the institutional level. These reforms would formally recognize collaborative accomplishments, such as interdisciplinary publications, joint grant awards, and measurable community impact, alongside traditional individual achievements. Furthermore, the creation of academic and professional pathways that explicitly value transdisciplinary work is essential to ensuring long-term cultural and structural change.

*Application of the Framework to Demographic Challenges*

The transdisciplinary framework can be applied systematically to address the distinct yet interconnected demographic challenges:

1. **Declining Birth Rates:** To tackle declining birth rates, a transdisciplinary approach would involve a diverse group of stakeholders, including demographers, economists, sociologists, public health experts (focusing on infertility and mental health), ethicists, urban planners, women's rights advocates, community leaders, and, critically, individuals and couples considering parenthood. Through co-creation, policies can be developed that holistically address both socio-economic factors, such as affordable housing, accessible childcare, and comprehensive parental leave, and socio-cultural factors, including evolving gender roles and career pressures, all while upholding reproductive freedom. This approach moves beyond simplistic pronatalist policies and toward more nuanced, rights-based strategies.

   The current focus on pronatalist policies often overlooks the deeper societal shifts driving fertility decline, such as women's increased human capital and career focus. A transdisciplinary approach would reveal that simply incentivizing births without addressing underlying structural inequalities or personal aspirations could lead to policy failure or even unintended negative consequences. For example, China's One Child Policy serves as a cautionary tale of how policies implemented without consideration of long-term systemic impacts can have profound negative effects.

Governments have implemented pronatalist policies and matchmaking events with the aim of increasing birth rates. However, well-educated and professional women often face higher opportunity costs and are less likely to desire more children. Working adults increasingly prioritize careers and personal development over immediate parenthood. These realities indicate a fundamental shift in societal values and individual priorities. If policies only address the symptom of a low birth rate through financial incentives or social pressure, without addressing root causes such as the high opportunity cost for women, the desire for personal development, or the financial burden of raising children in urban areas, they are unlikely to be effective. Worse, they may create social imbalances and undue pressure. A transdisciplinary approach is essential to understand the complex interplay of these factors and to co-create solutions that align with modern societal values and individual autonomy. This shifts the focus from merely trying to increase birth rates to creating a society where individuals can genuinely choose to have children, if they wish, and are fully supported in that choice.

2. **Aging Population and Healthcare:** For the aging population and healthcare, stakeholders would include geriatricians, nurses, social workers, physical therapists, urban planners, economists, policymakers, AI developers, older adults themselves, and their caregivers. Co-creation would focus on designing integrated care models that address both medical needs and the social determinants of health, such as income, housing, transportation, and social support. It would also involve developing AI solutions for geriatric care, such as remote monitoring and robotic assistance, that are co-designed with older adults to ensure usability, address privacy concerns, and incorporate humanistic values like compassion.

   The phenomenon of brain drain, in which healthcare professionals migrate from middle- and low-income countries to high-income countries, is not merely a national workforce issue. It represents a global equity challenge. High-income countries recruit healthcare professionals from less wealthy nations to address their own shortages. While this provides job opportunities and higher salaries for professionals from developing countries, it simultaneously creates a shortage of skilled professionals in their home countries, which worsens healthcare outcomes in those settings. A national solution, such as developed countries recruiting more professionals, risks perpetuating this imbalance.

   A transdisciplinary framework applied at an international level could involve global health organizations, economic development agencies, and national governments working together to co-create sustainable solutions. Such collaboration would address the needs of aging populations in high-income countries while also strengthening healthcare infrastructure in lower-income regions. Possible actions include investing in training and infrastructure in source countries, developing ethical recruitment guidelines, and establishing models of shared care that provide mutual benefit.

3. **AI Integration in Healthcare:** The integration of AI in healthcare requires stakeholders such as AI developers, ethicists, legal experts, clinicians, patients (especially older adults), policymakers, and social scientists. Through co-creation, AI systems can be developed that are not only technologically advanced but also ethically sound, user-friendly, and compassionate.

This would involve a team of AI developers, ethicists, clinicians, and older adults working together to refine models. This process ensures that data scientists receive continuous feedback from clinicians to improve algorithmic accuracy, while clinicians learn the capabilities and limitations of the AI tools, effectively bridging the professional culture gap. This iterative loop is essential for building trustworthy systems that can be safely integrated into clinical workflows (Hinostroza Fuentes et al., 2025; Tan et al., 2025b).

This involves integrating the human-AI system of intelligent caring framework from the design phase, ensuring robust data privacy compliance, and building trust through rigorous validation and transparent communication. Furthermore, addressing cost and accessibility issues through public-private partnerships is crucial.

The reluctance of older adults to embrace change, combined with a lack of trust in digital medical devices, represents a critical barrier to AI adoption. This is not simply a technological problem; it is a complex socio-psychological challenge. Older adults have expressed privacy concerns regarding voice assistant systems due to potential risks of identity and financial theft. This highlights that the true barriers lie in trust, perceived risk, and discomfort with change, rather than in technical capability alone.

Even the most advanced AI will fail if it is not accepted and trusted by its intended users. A transdisciplinary approach would involve behavioral scientists, gerontologists, and user experience designers working directly with older adults to understand their concerns, co-design intuitive interfaces, and build trust through education and demonstrated reliability. This reframes the question from "what AI can do" to "how AI can be integrated into human lives effectively and ethically." It ensures that AI tools are not only functional but also trustworthy, empathetic, and culturally appropriate for their users.

**Table 1. Proposed Transdisciplinary Framework for Demographic Challenges.**

| Framework Principle / Component | Demographic Challenge Addressed | Key Stakeholders Involved | Example Actions/Outcomes |
|---|---|---|---|
| Shared Problem Definition | Declining Birth Rates | Demographers, Economists, Sociologists, Individuals/Families, Policymakers | Collaborative workshops to identify root causes beyond economic incentives. |
| Inclusive Stakeholder Engagement | Aging Population & Healthcare | Geriatricians, Social Workers, Older Adults, Caregivers, Urban Planners | Co-design of community-based care models and accessible urban environments. |
| Iterative Co-creation | AI Integration in Healthcare | AI Developers, Clinicians, Ethicists, Older Adults (users) | Agile development of AI tools with continuous user feedback for usability and trust. |
| Ethical Integration | AI Integration in Healthcare | Ethicists, Legal Experts, AI Developers, Patient Advocates | Development of AI systems adhering to data privacy and compassion principles (e.g., "human-AI system of intelligent caring"). |
| Capacity Building | All Challenges | Academics, Practitioners, Community Members | Cross-disciplinary training programs; "champions" bridging divides. |
| Flexible Governance | All Challenges | Project Leaders, Institutional Management, Policymakers | Policies rewarding team achievements; clear conflict resolution protocols. |

Table 1 provides a visual synthesis of the complex interactions and multi-stakeholder engagement that underpin the proposed framework. By mapping core principles to demographic challenges and the corresponding stakeholders, the table distills abstract concepts into actionable insights. This structured presentation reinforces the framework's holistic nature and enhances its practical utility as a quick-reference guide for readers. Given the inherent complexity of transdisciplinarity, encompassing guiding principles, systemic challenges, and diverse actors, a tabular format effectively organizes and communicates these interconnections. It demonstrates how academic and non-academic perspectives converge to address multifaceted demographic issues, while also suggesting clear entry points for implementation. In doing so, the table bridges theory and practice, making the framework more concrete, accessible, and applicable across contexts.

**Conclusion**

Developed nations are grappling with profound demographic shifts, most notably, declining birth rates driven by socio-economic, sociocultural, and health-related factors, alongside a rapidly aging population that places increasing strain on healthcare systems and labor markets. While governments have responded with various policy measures, addressing these complex challenges requires a comprehensive, transdisciplinary approach. Such a framework, which integrates insights from diverse academic disciplines and non-academic stakeholders, fosters innovation, improves care quality, and supports the development of ethical, contextually appropriate, and impactful solutions.

To move beyond fragmented interventions, both policy and practice must adopt holistic, transdisciplinary strategies. This includes reimagining healthcare delivery through integrated, team-based models; establishing strong ethical frameworks for the governance of AI technologies; promoting international collaboration to address workforce imbalances; and prioritizing trust-building and community engagement to support successful implementation.

Future research should empirically assess the effectiveness of transdisciplinary frameworks, identify strategies to overcome barriers to collaboration, and pursue longitudinal studies focused on the development and ethical deployment of AI in healthcare. Further investigation is also warranted into economic models for integrated demographic solutions, the sociocultural underpinnings of population change, and the broader dynamics of global health equity and migration.